\documentclass[10pt,a4paper]{article}
\usepackage{amsmath}
\usepackage{amssymb}
\usepackage{amsfonts}
\usepackage{amstext}
\usepackage{amsbsy}
\usepackage[mathscr]{eucal}
\usepackage{graphicx}
\usepackage{color}
\usepackage[all]{xy}
\newcommand{\beq}{\begin{equation}}
\newcommand{\eeq}{\end{equation}}
\newcommand{\beqa}{\begin{eqnarray}}
\newcommand{\eeqa}{\end{eqnarray}}
\newcommand{\ket}[1]{| #1 \rangle}

 \numberwithin{equation}{subsection}

\makeindex
\title{\Large\textbf{Geometrical structure of entangled states and secant variety}}

\author{\textit{ Hoshang Heydari}\\
        \small\textit{Institute of Quantum
Science, Nihon University,}\\
\small\textit{1-8 Kanda-Surugadai, Chiyoda-ku, Tokyo 101-8308,
Japan}
\\\small\textit{Email: hoshang@edu.cst.nihon-u.ac.jp}}

\date{}
\begin{document}
\maketitle \thispagestyle{empty} \maketitle 

\begin{abstract}
We show that the secant variety of the Segre variety gives useful
information about the geometrical structure of an arbitrary
multipartite quantum system. In particular, we investigate the
relation between arbitrary bipartite and three-partite entangled
states and this secant variety. We also discuss the geometry of an
arbitrary general multipartite state.
\end{abstract}
\section{Introduction}
Recently, the geometry and topology of entanglement has got more
attention and we know more about  the geometrical structure of pure
multipartite entangled quantum states. We have also  managed to
construct some useful measures of entanglement based on these
underlying geometrical structures. However, we know less about the
geometrical structure of an arbitrary multipartite quantum state and
there is a need for further investigation on these states.
Concurrence is a measure of entanglement
 which is directly related to
the entanglement of formation \cite{Wootters98}. Its geometrical
structure is hidden in a map called Segre embedding
\cite{Dorje99,Miyake,Hosh1, Hosh2}. The Segre variety is generated
by the quadratic polynomials that correspond to the separable set of
pure multipartite states. We can construct a measure of entanglement
for bipartite and three-partite states based on the Segre variety
\cite{Hosh1}. We can also construct a measure of entanglement for
general pure multipartite states based on a modification of the
Segre variety by adding similar quadratic polynomials \cite{Hosh2}.
In this paper, we will establish a relation between the secant
variety of the Segre variety  and multipartite states. For example
we show that the concurrence of arbitrary bipartite and
three-partite entangled states are equivalent to the secant variety
of the Segre variety. We also generalize our result for a measure of
entanglement for an arbitrary multipartite state.
In section \ref{com}, we will  define the complex projective
variety. We also introduce the Segre embedding and the Segre variety
for general pure multipartite states. In following section
\ref{sec}, we will define and discuss the secant variety of a
projective variety. In section \ref{secseg}, we investigate the
secant variety of the Segre variety, which is of central importance
in this paper. Moreover, we investigate relation and relevance of
the secant variety of the Segre variety as geometrical structure of
entangled and separable states. For example, we show that this
variety defines the space of the concurrence of a mixed state.
Finally, in section \ref{secsegmulti}, we expand our result to an
arbitrary multipartite state.
 As usual, we denote a general,
composite quantum system with $m$ subsystems as
$\mathcal{Q}=\mathcal{Q}_{m}(N_{1},N_{2},\ldots,N_{m})
=\mathcal{Q}_{1}\mathcal{Q}_{2}\cdots\mathcal{Q}_{m}$, consisting of
the pure states $
\ket{\Psi}=\sum^{N_{1}}_{k_{1}=1}\sum^{N_{2}}_{k_{2}=1}\cdots\sum^{N_{m}}_{k_{m}=1}
$ $\alpha_{k_{1},k_{2},\ldots,k_{m}} \ket{k_{1},k_{2},\ldots,k_{m}}
$ and corresponding to the Hilbert space $
\mathcal{H}_{\mathcal{Q}}=\mathcal{H}_{\mathcal{Q}_{1}}\otimes
\mathcal{H}_{\mathcal{Q}_{2}}\otimes\cdots\otimes\mathcal{H}_{\mathcal{Q}_{m}}
$, where the dimension of the $j$th Hilbert space is
$N_{j}=\dim(\mathcal{H}_{\mathcal{Q}_{j}})$. We are going to use
this notation throughout this paper. In particular, we denote a pure
two-qubit state by $\mathcal{Q}^{p}_{2}(2,2)$.
 Concurrence is a widely used measure of entanglement which have
 been successfully applied to many different field of finite quantum
 systems. It also gives an analytical expression for entanglement of
 formation for bipartite quantum systems.
In Wootter's definition of concurrence for a two-qubit state
\cite{Wootters98}, the tilde
 operation is an example of conjugation, i.e., an antiunitary
 operator. Based on this observation, Uhlmann \cite{Uhlmann00}
 generalized the concept of concurrence. Uhlmann considered an
 arbitrary conjugation acting on an arbitrary Hilbert space. For
 example
 concurrence for a pure quantum system
$\mathcal{Q}^{p}_{2}(N_{1},N_{2})$  is defined by
\begin{eqnarray}
\mathcal{C}
\left(\mathcal{Q}^{p}_{2}(N_{1},N_{2})\right)=\sqrt{|\langle\Psi\ket{\Theta\Psi}|},
\end{eqnarray}
where $\Theta$ is an antilinear operator and satisfies
$\Theta^{2}=I$. Moreover, the
 concurrence of a quantum system
$\mathcal{Q}_{2}(N_{1},N_{2})$ is defined by
\begin{eqnarray}
\mathcal{C}
\left(\mathcal{Q}_{2}(N_{1},N_{2})\right)=\inf\sum_{i}p_{i}\mathcal{C}^{i}
\left(\mathcal{Q}^{p}_{2}(N_{1},N_{2})\right),
\end{eqnarray}
where the infimum is taken over all pure state decompositions. If we
have a quantum system $\mathcal{Q}_{2}(2,2)$, where $\Theta$ is a
tilde operation, then concurrence coincides with Wootter's formula
for concurrence of two-qubit states.

\section{Complex projective variety}\label{com}
 In this section, we review some basic definition of complex projective variety. The general reference on projective algebraic geometry can be
found in \cite{Griff78,Hart77,Mum76}.
Let $C[z]=C[z_{1},z_{2}, \ldots,z_{n}]$ denote the polynomial
algebra in $n$  variables with complex coefficients. Then, given a
set of $q$ polynomials $\{h_{1},h_{2},\ldots,h_{q}\}$ with $h_{i}\in
C[z]$, we define a complex affine variety as
\begin{eqnarray}
&&\mathcal{V}_{\mathbb{C}}(h_{1},h_{2},\ldots,h_{q})=\{P\in\mathbb{C}^{n}:
h_{i}(P)=0~\forall~1\leq i\leq q\},
\end{eqnarray}
where $P=(a_{1},a_{2}, \ldots,a_{n})$ is called a point of
$\mathbb{C}^{n}$ and the $a_{i}$ are called the coordinates of $P$.
A complex projective space $\mathbb{P}^{n}$ is defined to be the set
of lines through the origin in $\mathbb{C}^{n+1}$, that is, $
\mathbb{P}^{n}=\mathbb{C}^{n+1}-{0}/ \sim$, where the equivalence
relation $\sim$ is defined as follow;
$(x_{1},\ldots,x_{n+1})\sim(y_{1},\ldots,y_{n+1})$ for $\lambda\in
\mathbb{C}-0$, where $y_{i}=\lambda x_{i}$ for all $ 0\leq i\leq
n+1$. Given a set of homogeneous polynomials
$\{h_{1},h_{2},\ldots,h_{q}\}$  with $h_{i}\in C[z]$, we define a
complex projective variety as
\begin{eqnarray}
&&\mathcal{V}(h_{1},\ldots,h_{q})=\{O\in\mathbb{P}^{n}:
h_{i}(O)=0~\forall~1\leq i\leq q\},
\end{eqnarray}
where $O=[a_{1},a_{2},\ldots,a_{n+1}]$ denotes the equivalent class
of points $\{\alpha_{1},\alpha_{2},\ldots,$
$\alpha_{n+1}\}\in\mathbb{C}^{n+1}$. We can view the affine complex
variety
$\mathcal{V}_{\mathbb{C}}(h_{1},h_{2},\ldots,h_{q})\subset\mathbb{C}^{n+1}$
as a complex cone over the complex projective variety
$\mathcal{V}(h_{1},h_{2},\ldots,h_{q})$.

As an important example of projective variety we will discuss the
Segre variety. For a multipartite quantum system
$\mathcal{Q}(N_{1},\ldots,N_{m})$, let
$\overline{N}=(N_{1}-1,\ldots,N_{m}-1)$ and
$V_{1},V_{2},\ldots,V_{m}$ be vector spaces over the field of
complex numbers $\mathbb{C}$, where $\dim V_{j}=N_{j}$. That is, we
have $\mathbb{P}^{N_{j}-1}=\mathbb{P}(V_{j})$ for all $j$. Then we
define a Segre map by
\begin{equation}
\mathcal{S}_{N_{1},N_{2},\ldots,N_{m}}:\mathbb{P}^{N_{1}-1}\times\mathbb{P}^{N_{2}-1}\times\cdots\times\mathbb{P}^{N_{m}-1}\longrightarrow\mathbb{P}^{\mathcal{N}-1},
\end{equation}
where $\mathcal{N}=\prod^{m}_{j=1}N_{j}$. This map is based on the
canonical multilinear map
\begin{equation}
\begin{array}{ccc}
  V_{1}\times V_{2}\times\cdots\times V_{m} & \rightarrow &V_{1}\otimes V_{2}\otimes\cdots\otimes V_{m}\\
  v_{1}\times v_{2}\times\cdots\times v_{m} & \mapsto &v_{1}\otimes v_{2}\otimes\cdots\otimes v_{m}\\
\end{array}
\end{equation}
Thus, we have
$\mathbb{P}^{\mathcal{N}-1}=\mathbb{P}(V_{1},\ldots,V_{m})$. The
Segre variety
$\mathfrak{S}_{\overline{N}}=\mathrm{Im}(\mathcal{S}_{N_{1},N_{2},\ldots,N_{m}})$
is defined to be the image of the Segre embedding. By definition,
the Segre variety  is formed by the set of all classes of
decomposable tensors in $\mathbb{P}^{\mathcal{N}-1}$. For  a quantum
system $\mathcal{Q}(N_{1},\ldots,N_{m})$, the Segre variety is given
by
\begin{eqnarray}\label{eq: submeasure}
\mathfrak{S}_{\overline{N}}&=&\bigcap_{\forall
j}\mathcal{V}(\alpha_{k_{1},k_{2},\ldots,k_{m}}\alpha_{l_{1},l_{2},\ldots,l_{m}}\\\nonumber&&-
\alpha_{k_{1},k_{2},\ldots,k_{j-1},l_{j},k_{j+1},\ldots,k_{m}}\alpha_{l_{1},l_{2},
\ldots,l_{j-1},k_{j},l_{j+1},\ldots,l_{m}}).
\end{eqnarray}
We can also partition the Segre embedding as follows:
$$\xymatrix{ (\mathbb{P}^{N_{1}-1}\times\cdots\times\mathbb{P}^{N_{l}-1})
\times(\mathbb{P}^{N_{l}-1}\times\cdots\times\mathbb{P}^{N_{m}-1}
\ar[d]_{\mathcal{S}_{N_{1},N_{2},\ldots,N_{m}}})\ar[r]&\mathbb{P}^{\mathcal{M}_{1}}
\times\mathbb{P}^{\mathcal{M}_{2}}\ar[d]_{I\otimes I}\\
             \mathbb{P}^{N_{1}N_{2}\cdots N_{m}-1}&\mathbb{P}^{\mathcal{M}_{1}}\times\mathbb{P}^{\mathcal{M}_{2}}
             \ar[l]_{\mathcal{S}_{\mathcal{M}_{1},\mathcal{M}_{2}}}}$$
where $\mathcal{M}_{1}=N_{1}N_{2}\ldots
N_{l}-1$,$\mathcal{M}_{2}=N_{l+1}N_{l+2}\ldots N_{m}-1$ and
$(\mathcal{M}_{1}+1)(\mathcal{M}_{2}+1)=N_{1}N_{2}\ldots N_{m}$. For
the Segre variety, which is represented by a completely decomposable
tensors, the above diagram commutate.
 Let $\mathbb{X}\subset \mathbb{P}^{N}$. Then there are two important
subsets of $ \mathbb{P}^{N}$; the secant variety
$\mathfrak{Sec}(\mathbb{X})$, which is defined to be the closure of
the set of point lying on secant $\overline{x_{1}x_{2}}$, where
$x_{1}$ and $x_{2}$ are distinct points of $\mathbb{X}$. The second
one $\mathfrak{Tan}(\mathbb{X})$ is the union of the projective
tangent spaces. In the next section we will discuss the secant
variety of a projective variety.
\section{Secant variety}\label{sec}
The secant variety of a projective variety has been  studied in
algebraic geometry and some recent references include
\cite{Catal1,Catal2}.
 The $k$-th secant
variety $\mathfrak{Sec}_{k}(\mathbb{X})$ of
$\mathbb{X}\subset\mathbb{P}^{M}$ with $\dim \mathbb{X}=d$ is
defined to be the closure of the union of $k$-dimensional linear
subspaces of $\mathbb{P}^{M}$ determined by general $k+1$ points on
$\mathbb{X}$
\begin{equation}
\mathfrak{Sec}_{k}(\mathbb{X})=\overline{\bigcup\{\text{all
secant}~\mathbb{P}^{k}~\text{'s to}~\mathbb{X}\}},
\end{equation}
where for $P_{0},P_{1}\ldots P_{k}\in \mathbb{X}$, we have
$\mathbb{P}^{k}=\langle P_{0},P_{1}\ldots P_{k}\rangle$.
 Moreover, the dimension of $\mathfrak{Sec}_{k}(\mathbb{X})$ satisfies
\begin{equation} \dim\mathfrak{Sec}_{k}(\mathbb{X})\leq
\min\{M^{'},(k+1)(d+1)-1\},
\end{equation}
where $M^{'}$ is the dimension of the linear subspace spanned by
$\mathbb{X}$. The subvariety $\mathbb{X}$ is called $k$-defect when
$\dim\mathfrak{Sec}_{k}(\mathbb{X})<\min\{M^{'},(k+1)(d+1)-1\}$. For
example, the secant variety of Segre variety
$\mathfrak{Sec}_{k}(\mathfrak{S}_{\overline{N}})$ is the closure of
the set of classes of those tensor products which can be written as
the sum of at most $k+1$ decomposable tensor products. Thus, the
secant variety of Segre variety
$\mathfrak{Sec}_{k}(\mathfrak{S}_{\overline{N}})$ gives some useful
information about the geometry of entangled and separable  mixed
multipartite states.

\section{Secant variety of the Segre variety and concurrence}\label{secseg}
In this section,  we investigate the secant variety of the Segre
variety and show that the geometry of  concurrence for mixed
bipartite and three-partite states is given by this variety. In the
next section, we will discuss the secant variety of variety for an
arbitrary multipartite state. For bipartite quantum system
$\mathcal{Q}(N_{1},N_{2})$, the Segre variety
$\mathfrak{S}_{\overline{N}}$ is the variety of $N_{1}\times N_{2}$
matrices of rank 1. Thus the secant variety
$\mathfrak{Sec}_{k}(\mathfrak{S}_{N_{1},N_{2}})$ is the matrices of
rank less than $k$ and $k=N_{1}$ is the least integer for which
$\mathfrak{Sec}_{k}(\mathfrak{S}_{N_{1},N_{2}})=\mathbb{P}^{N_{1}N_{2}-1}$.
The Segre variety has two rulings by the families of linear spaces
$v\otimes\mathbb{P}(W) $ and $\mathbb{P}(V)\otimes w$ for all $v\in
V$ and $w\in W$.  The Segre variety can be seen as decomposable
tensors in $ \mathbb{P}(V)\otimes \mathbb{P}(W)$. The $k$-fold
secant plane to the Segre variety is given by the tensor of rank
$k$. For example, a tensor which can be written as $ \sum^{k}_{i=1}
v_{i}\otimes w_{i}=v_{1}\otimes w_{1}+v_{2}\otimes
w_{2}+\cdots+v_{k}\otimes w_{k}.$
As an example, we will discuss the Secant variety of the  Segre
variety $\mathfrak{Sec}_{k}(\mathfrak{S}_{(3,3)})$ of quantum system
$\mathcal{Q}_{2}(3,3)$. For this quantum system, the Segre variety
is given by $\mathfrak{S}_{(3,3)}=\bigcap^{3}_{
k_{1},l_{1},k_{2},l_{2}=1}\mathcal{V}(\alpha_{k_{1},k_{2}}\alpha_{l_{1},l_{2}}-\alpha_{k_{1},l_{2}}\alpha_{l_{1},k_{2}})$.
Moreover, we have
$\dim\mathfrak{Sec}_{1}(\mathfrak{S}_{(3,3)})=(3+3)(1+1)-(1+1)^{2}-1=7$,
but the expected dimension was
$\dim\mathfrak{Sec}_{1}(\mathfrak{S}_{(3,3)})\leq
\min\{M^{'},(k+1)(d+1)-1\}=\min\{8,(1+1)(4+1)-1\}=8$. We have
expected that the secant variety
$\mathfrak{Sec}_{1}(\mathfrak{S}_{(3,3)})$  does fill the enveloping
space and this is an example of a deficient Segre variety. For
bipartite systems, if we assume that $N_{1}<N_{2}$, then for all
$1\leq k<N_{1}$ the secant variety
$\mathfrak{Sec}_{k}(\mathfrak{S}_{(N_{1},N_{2})})$ has dimension
less than the expected dimension and the least $k$ for which
$\mathfrak{Sec}_{k}(\mathfrak{S}_{(N_{1},N_{2})})$ fills its
enveloping space is $k=N_{1}$.
 Next, we write the
 concurrence of a quantum system
$\mathcal{Q}_{2}(N_{1},N_{2})$ as follows
\begin{eqnarray}
\mathcal{C}
\left(\mathcal{Q}_{2}(N_{1},N_{2})\right)&=&\inf\sum_{i}p_{i}\mathcal{C}
\left(\Psi_{i}\right)\\\nonumber&=&\inf\sum_{i}p_{i}
\left(\mathcal{N}\sum^{N_{1}}_{k_{1},l_{1}=1}\sum^{N_{2}}_{k_{2},l_{2}=1}
\left|\alpha^{i}_{k_{1},k_{2}}\alpha^{i}_{l_{1},l_{2}}-
\alpha^{i}_{k_{1},l_{2}}\alpha^{i}_{l_{1},k_{2}}\right|^{2}\right)^{\frac{1}{2}}
\\\nonumber&\simeq&\inf
\left(\sim\text{sum of all decom. ten. in}
~\mathbb{P}^{N_{1}N_{2}-1} \right)
\\\nonumber&\simeq&\inf
\mathfrak{Sec}_{k}(\mathfrak{S}_{N_{1},N_{2}}),
\end{eqnarray}
where $N$ is a normalization constant. From this expression we can
see that the geometry of concurrence of arbitrary  bipartite state
is given by the secant variety of the Segre variety. Moreover, the
geometry of arbitrary three-partite states can be given by this
secant variety, since we can construct a measure for three-partite
states based on the Segre variety in the same way as we did  for
bipartite states. The generalized concurrence  for such a state is
given by
\begin{eqnarray}
\mathcal{C}(\mathcal{Q}^{p}_{3}(N_{1},N_{2},N_{3}))&=&(\mathcal{N}\sum^{m=3}_{k_{1},l_{1};k_{2},l_{2};k_{3},l_{3}}
\sum_{\forall
j}|\alpha_{k_{1},k_{2},\ldots,k_{m}}\alpha_{l_{1},l_{2},\ldots,l_{m}}\\\nonumber&&-
\alpha_{k_{1},k_{2},\ldots,k_{j-1},l_{j},k_{j+1},\ldots,k_{m}}\alpha_{l_{1},l_{2},
\ldots,l_{j-1},k_{j},l_{j+1},\ldots,l_{m}}|^{2})^{\frac{1}{2}}.
\end{eqnarray}
From this equation and the discussion about the concurrence of
bipartite states, we have
\begin{eqnarray}\nonumber
\mathcal{C}(\mathcal{Q}_{3}(N_{1},N_{2},N_{3}))&=&\inf\sum_{i}p_{i}\mathcal{C}^{i}(\mathcal{Q}^{p}_{3}(N_{1},N_{2},N_{3}))
\\\nonumber&\simeq&\inf
\mathfrak{Sec}_{k}(\mathfrak{S}_{N_{1},N_{2},N_{3}}).
\end{eqnarray}
 We can also connect the secant variety of the
Segre variety to the separable set of multipartite states
$\mathcal{Q}(N_{1},N_{2},\ldots,N_{m})$ based on the relation
between perfect codes and Secant variety of the Segre variety. The
existence of perfect codes can be proved based on finite fields with
$q$ elements. The perfect code  exist only for the following
parameters: $q$ is a prime power, $t=\frac{q^{l}-1}{q-1}$ for $l\geq
2$ and $k=q^{t-l}$. Let us look at some examples of this kind. Let
$q=2$, $t=2^{l}-1$, and $k=q^{t-l}$, where $l$ is a positive number.
Then for the Segre embedding
\begin{equation}
\mathcal{S}_{2,2,\ldots,2}:\overbrace{\mathbb{P}^{1}\times\mathbb{P}^{1}\times\cdots
\times\mathbb{P}^{1}}^{t-\text{times}}\longrightarrow\mathbb{P}^{2^{t}-1},
\end{equation}
the secant variety of the corresponding Segre variety
$\mathfrak{Sec}_{k-1}(\mathfrak{S}_{t})=\mathbb{P}^{2^{t}-1}$ which
fits exactly  into its enveloping space.  Thus, all
$\mathfrak{Sec}_{k-1}(\mathfrak{S}_{t})$ have the expected
dimension. This secant variety coincide with the space of separable
mixed multi-qubits states $\mathcal{Q}(2,2,\ldots,2)$.

Next, let $q$ be a prime power. Then for any $l\geq1$,
$t=\frac{q^{l}-1}{q-1}$,  and the Segre embedding
\begin{equation}
\mathcal{S}_{q,q,\ldots,q}:\overbrace{\mathbb{P}^{q-1}\times\mathbb{P}^{q-1}\times\cdots
\times\mathbb{P}^{q-1}}^{t-\text{times}}\longrightarrow\mathbb{P}^{q^{t}-1},
\end{equation}
the secant variety of the Segre variety
$\mathfrak{Sec}_{k-1}(\mathfrak{S}_{t})=\mathbb{P}^{q^{t}-1}$ gives
information on the geometry  of the entangled and separable sets of
an arbitrary quantum system $\mathcal{Q}(q,q,\ldots,q)$.

\section{Secant variety and arbitrary general multipartite state}\label{secsegmulti}

Recently, we have proposed a measure of entanglement for general
pure multipartite states as \cite{Hosh2}
\begin{eqnarray}\label{EntangSeg2}\nonumber
&&\mathcal{F}(\mathcal{Q}^{p}_{m}(N_{1},\ldots,N_{m}))
=(\mathcal{N}\sum_{\forall \sigma\in\text{Perm} (u)}\sum_{
k_{j},l_{j}, j=1,2,\ldots,m}\\&&|\alpha_{k_{1}k_{2}\ldots
k_{m}}\alpha_{l_{1}l_{2}\ldots l_{m}} -
\alpha_{\sigma(k_{1})\sigma(k_{2})\ldots\sigma(k_{m})}\alpha_{\sigma(l_{1})\sigma(l_{2})
\ldots\sigma(l_{m})}|^{2})^{\frac{1}{2}},
\end{eqnarray}
where $\sigma\in\text{Perm}(u)$ denotes all possible sets of
permutations of indices for which $k_{1}k_{2}\ldots k_{m}$ are
replace by $l_{1}l_{2}\ldots l_{m}$, and $u$ is the number of
indices to permute. By construction this measure of entanglement
vanishes on product states and it is also invariant under all
possible permutations of indices. Note that the first set of
permutations defines the Segre variety, but there are also
additional complex projective variety embedded in
$\mathbf{CP}^{\mathcal{N}-1}$ which are defined by other sets of
permutations of indices in equation (\ref{EntangSeg2}). We can also
apply the same procedure as in the case of the concurrence
 to define a measure of entanglement for arbitrary multipartite
 states
\begin{eqnarray}\label{EntangSeg2}\nonumber
&&\mathcal{F}(\mathcal{Q}_{m}(N_{1},\ldots,N_{m}))=\inf\sum_{i}p_{i}\mathcal{F}^{i}(\mathcal{Q}^{p}_{m}(N_{1},\ldots,N_{m}))
\\\nonumber&&=\inf\sum_{i}p_{i}(\mathcal{N}\sum_{\forall \sigma\in\text{Perm}
(u)}\sum_{ k_{j},l_{j},
j=1,2,\ldots,m}\\&&|\alpha^{i}_{k_{1}k_{2}\ldots
k_{m}}\alpha^{i}_{l_{1}l_{2}\ldots l_{m}} -
\alpha^{i}_{\sigma(k_{1})\sigma(k_{2})\ldots\sigma(k_{m})}\alpha^{i}_{\sigma(l_{1})\sigma(l_{2})
\ldots\sigma(l_{m})}|^{2})^{\frac{1}{2}}.
\end{eqnarray}
Next, for  a quantum system $\mathcal{Q}(N_{1},\ldots,N_{m})$ we
define the variety
\begin{eqnarray}\label{eq: submeasure}
\mathfrak{T}_{\overline{N}}&=&\bigcap_{\forall \sigma\in\text{Perm}
(u), k_{j},l_{j},
j=1,2,\ldots,m}\mathcal{V}(\alpha_{k_{1},k_{2},\ldots,k_{m}}\alpha_{l_{1},l_{2},\ldots,l_{m}}\\\nonumber&&-
\alpha_{\sigma(k_{1})\sigma(k_{2})\ldots\sigma(k_{m})}\alpha_{\sigma(l_{1})\sigma(l_{2})
\ldots\sigma(l_{m})}).
\end{eqnarray}
which  include the Segre variety. Now, based on our discussion about
the concurrence of bipartite and three-partite states, we conclude
that
\begin{eqnarray}\label{EntangSeg2}\nonumber
\mathcal{F}(\mathcal{Q}_{m}(N_{1},\ldots,N_{m}))&\simeq&\inf
\mathfrak{Sec}_{k}(\mathfrak{T}_{\overline{N}}).
\end{eqnarray}
This equivalence relation establish a relationes between  the
geometrical structure of a measure of entanglement for arbitrary
general multipartite states and the secant variety
$\mathfrak{Sec}_{k}(\mathfrak{T}_{\overline{N}})$.

We have established a connection between pure mathematics and
fundamental quantum mechanics with some applications in the field of
quantum information and computing. We have introduced and discussed
the secant variety of  the Segre variety. But the secant varieties
are still subject of research in algebraic geometry. For example,
there are still many fundamental open questions about the secant
variety of the Segre variety.  However, we hope that this
geometrical structure may give us some hint to how to solve the
problem of quantifying entanglement of an arbitrary multipartite
system.
\begin{flushleft}
\textbf{Acknowledgments:} The  author gratefully acknowledges the
financial support of the Japan Society for the Promotion of Science
(JSPS).
\end{flushleft}


\end{document}